\documentclass[letterpaper, 10 pt, conference]{ieeeconf}  

\usepackage{graphicx}
\usepackage{amsmath,amssymb,amsfonts}
\usepackage{multirow}
\usepackage{psfrag}
\usepackage{subfigure}
\usepackage{color}
\usepackage{soul}
\usepackage{xcolor}
\usepackage[hyphens]{url}
\usepackage[bookmarks=false]{hyperref}
\usepackage[hyphenbreaks]{breakurl}
\usepackage{lipsum}
\usepackage[ruled,vlined,linesnumbered]{algorithm2e}
\usepackage{mathrsfs}
\usepackage{pbox}
\usepackage{wrapfig}
\usepackage{array}

\DontPrintSemicolon

\newcommand{\matr}[1]{\mathbf{#1}}

\DeclareMathOperator*{\argminA}{arg\,min}

\makeatletter
\def\hlinewd#1{%
  \noalign{\ifnum0=`}\fi\hrule \@height #1 \futurelet
   \reserved@a\@xhline}
\makeatother

\graphicspath{{Figures/}}

\IEEEoverridecommandlockouts                              

\title{\LARGE \bf
Towards Fully Autonomous Drone-Based Last-Mile Delivery
}

\author{Navid Mohammad Imran$^{1}$, Sabya Mishra$^{2}$, and Myounggyu Won$^{1}$
\thanks{$^{1}$Navid Mohammad Imran and Myounggyu Won are with the Department of Computer Science, University of Memphis, Memphis, TN, United States
        {\tt\small \{nimran, mwon\}@memphis.edu}}%
\thanks{$^{2}$Sabya Mishra is with the Department of Civil Engineering, University of Memphis, TN, United States
        {\tt\small smishra3@memphis.edu}}%
}

\begin{document}

\maketitle
\thispagestyle{empty}
\pagestyle{empty}

\begin{abstract}
The drone-based last-mile delivery is an emerging technology to deliver parcels using drones loaded on a truck. As more and more autonomous vehicles (AVs) will be available for delivery services, an opportunity is arising to fully automate drone-based last-mile delivery. In this paper, we integrate AVs with drone-based last-mile delivery aiming to fully automate the last-mile delivery process. We define a new problem called the autonomous vehicle routing problem with drones (A-VRPD). A-VRPD is to select AVs from a pool of available AVs and to schedule them to serve customers with an objective of minimizing the total operational cost. We formulate A-VRPD as an Integer Linear Programming (ILP) and propose a greedy algorithm to solve the problem based on real-world operational costs for different types of AVs, traveling distances calculated considering the current traffic conditions, and varying load capacities of AVs. Extensive simulations performed under various random delivery scenarios demonstrate that the proposed algorithm effectively increases profits for both the delivery company and AV owners compared with traditional VRP-D (and TSP-D) algorithm-based approaches.
\end{abstract}


\section{Introduction}
\label{sec:introduction}

The drone-based last-mile delivery~\cite{kitjacharoenchai2020two} is an emerging delivery solution that is increasingly adopted by many logistics companies like Amazon~\cite{amazon}, Google~\cite{google}, DHL~\cite{dhl}, and Alibaba~\cite{alibaba}. Basically, the idea is to use a truck loaded with one or more drones for delivery of parcels; a truck travels to a designated delivery location, and the drones are used to deliver parcels directly to customers. This is an ideal solution that maximizes the fast mobility and high accessibility of drones while addressing the problems of the limited load capacity and the constrained battery of drones by coupling with a truck.

Since Murray and Chu introduced a pioneering work on effectively routing and scheduling a drone and a truck to minimize the time required to serve all customer locations~\cite{murray2015flying}, numerous approaches have been proposed to make drone-based last-mile delivery more efficient~\cite{gonzalez2020truck}. Many of these approaches were focused on a single-drone-and-singe-truck scenario (the traveling salesman problem with a drone, TSP-D)~\cite{agatz2018optimization}\cite{dell2019drone}\cite{ha2018min}\cite{daknama2017vehicle}\cite{schermer2018algorithms}\cite{bouman2018dynamic}\cite{tang2019study}. Other works considered more general cases with multiple drones and multiple trucks (the vehicle routing problem with drones, VRP-D)~\cite{wang2017vehicle}\cite{poikonen2017vehicle}\cite{schermer2019matheuristic}\cite{sacramento2019adaptive}\cite{kitjacharoenchai2019multiple}. There are also solutions designed to address the variants of the problem such as minimizing the operational cost ~\cite{wang2017vehicle}\cite{poikonen2017vehicle}\cite{sacramento2019adaptive} and the energy consumption of drones~\cite{jeong2019truck}. We have presented the details of these various methods for drone-based last-mile delivery in Section~\ref{sec:related_work}.

One common assumption of existing solutions for drone-based last-mile delivery is that there are a limited number of human-driven trucks initially located at the package distribution center. Also, few solutions account for the current traffic conditions to compute the routes of trucks. On the other hand, due to the significant advancement of autonomous vehicle (AV) technology~\cite{duarte2018impact}, it is anticipated that a large number of AVs will be available for different kinds of mobility services~\cite{thomas2018reinventing}\cite{stocker2019shared}, \emph{e.g.,} as a taxi~\cite{narayanan2020shared} and for delivery of parcels~\cite{schlenther2020potential}. Adopting AVs for facilitating mobility services is expected to bring new challenges to traditional drone-based last-mile delivery systems.

In this paper, we define and investigate a novel autonomous vehicle routing problem with drones (A-VRPD) to select AVs from a pool of available AVs and schedule them to deliver parcels to customer locations to maximize the profits for both the logistics company and AV owners. \emph{A-VRPD is different from existing approaches for drone-based last-mile delivery in that it is focused on how to effectively utilize available AVs to optimize profits while existing works are concentrated on maximizing profits based on a fixed number of human-driven trucks.} We formulate A-VRPD as an integer linear programming (ILP) and propose a greedy heuristic algorithm to select AVs and to produce an optimal mapping between selected AVs and customer locations to maximize profits. More specifically, a tree-based cost-computation algorithm is designed to maximize profits based on real-world operational costs for different types of AVs, expected traveling distances and times calculated using Goole Map API (\emph{i.e.,} taking into account the current traffic conditions), varying load capacities of AVs, and available operation times of AVs.

Extensive simulations are performed to evaluate the effectiveness of the proposed algorithm. Numerous random delivery scenarios and varying numbers of available AVs are considered to validate the performance, in comparison with both the TSP-D and VRP-D algorithm-based approaches. The results demonstrate that the proposed greedy algorithm produces higher profits for both the delivery company and AV owners by 14.4\% (12.2\%) and 7.5\% (6.5\%), respectively, compared with those for an existing TSP-D algorithm-based (VRP-D algorithm-based) approaches. The results also show that the greedy algorithm reduces the delivery finish time by 6.3\% (2.3\%) compared with the TSP-D algorithm-based (VRP-D algorithm-based) approaches. The contributions of this paper are summarized as follows.

\begin{itemize}
  \item To the best of our knowledge, this is the first paper that defines the autonomous vehicle routing problem with drones (A-VRPD) to incorporate a pool of available AVs in the drone-based last-mile delivery, \emph{i.e.,} optimally selecting participating AVs and assigning them to customer locations to maximize profits for both the logistics company and AV owners.
  \item We formulate A-VRPD as an ILP and propose a greedy heuristic solution that takes into account the current traffic conditions.
  \item We perform extensive simulations under various random delivery scenarios and demonstrate that the proposed algorithm produces higher profits with faster delivery finish time compared with a traditional TSP/VRP-D algorithm-based methods.
\end{itemize}

In Section~\ref{sec:related_work}, we review the related work concentrating on drone-based last-mile delivery. The system model and notations used in this paper are presented in Section~\ref{sec:system_model}. We then formulate the A-VRPD problem as an ILP in Section~\ref{sec:problem_definition} and present the details of the greedy algorithm in Section~\ref{sec:greedy_approach}. The simulation results are analyzed in Section~\ref{sec:evaluation}, followed by the conclusion of this work in Section~\ref{sec:conclusion}.

\section{Related Work}
\label{sec:related_work}

This section presents a review on different approaches for drone-based last-mile delivery and their limitations. Since Murray and Chu introduced the problem of scheduling and routing a drone and a truck to minimize the time required for serving all customers and returning to the base station~\cite{murray2015flying}, various solutions have been proposed to enable drone-based last-mile delivery. The problem is called as the flying sidekick traveling salesman problem (FSTSP), or in some papers, the traveling salesman problem with a drone (TSP-D)~\cite{agatz2018optimization}.

\begin{table}[h]
\center
\label{table:related_work}
\caption{The max number of trucks and initial truck location}
\begin{tabular}{ p{3.7cm}|p{1.5cm}|p{1.5cm}}
  \hlinewd{1.5pt}
  References & Initial Truck Location & Max Number of Trucks \\ \hline
  \cite{yurek2021traveling,karak2019hybrid,murray2020multiple,wang2020cooperative,poikonen2019branch,jeong2019truck,poikonen2017vehicle,wang2017vehicle,tang2019study,ha2018min,gonzalez2020truck,murray2015flying,agatz2018optimization,dell2019drone} & Depot & 1 \\ \hline
  \cite{nguyen2021min,tamke2021branch,kitjacharoenchai2019multiple,sacramento2019adaptive,schermer2019matheuristic,bouman2018dynamic,schermer2018algorithms,daknama2017vehicle} & Depot & 2$\sim$5\\ \hline
  \cite{kitjacharoenchai2020two} & Depot & 14 \\ \hline
  \cite{gomez2021new} & Depot & 15 \\
  \hlinewd{1.5pt}
\end{tabular}
\end{table}

Various heuristic approaches have been developed to solve TSP-D. Schermer \emph{et al.} focused on the scalability of the problem and proposed two heuristic solutions, \emph{i.e.,} the two-phase heuristic (TPH) and single-phase heuristic (SPH)~\cite{schermer2018algorithms}. Bouman \emph{et al.} proposed a dynamic programming approach to solve TSP-D~\cite{bouman2018dynamic}. Tang \emph{et al.} created a constraint programming approach to solve the problem~\cite{tang2019study}. Poikonen \emph{et al.} developed four branch-and-bound-based heuristics~\cite{poikonen2019branch}. These approaches are, however, designed for a limited number of human-driven trucks (mostly a single truck), initially located at the distributed center (depot). Table~I summarizes the maximum number of trucks and their initial locations in existing approaches.

Variants of TSP-D have also been investigated. Ha \emph{et al.} addressed TSP-D focusing on the operational cost~\cite{ha2018min}. Jeong \emph{et al.} took into account the power consumption of drones and the restricted flying areas of the drones~\cite{jeong2019truck}. Dukkanci \emph{et al.}, similar to~\cite{ha2018min}, aimed to minimize the operational cost considering the energy consumption of drones~\cite{dukkanci2019drone}. Wang \emph{et al.} developed a multi-objective version of TSP-D to optimize both the operational cost and the time required to serve all customers~\cite{wang2020cooperative}. Similar to many solutions for TSP-D, most of these variants assume a limited number of human-driven trucks. Additionally, these approaches do not consider real-time traffic conditions in determining the routes of trucks.

Wang \emph{et al.} generalized TSP-D and introduced the vehicle routing problem with drones (VRP-D)~\cite{wang2017vehicle}. Compared to other solutions, multiple trucks and multiple drones are considered assuming that drones can be launched from a truck at any of the customer locations and the base station. Poikonen \emph{et al.} improved Wang's work by considering the limited battery of drones, varying distance metrics, operational cost in the objective function~\cite{poikonen2017vehicle}. Sacramento \emph{et al.} incorporated the time-limit constraint in the objective function and proposed an adaptive large neighborhood search metaheuristic~\cite{sacramento2019adaptive}. Schermer \emph{et al.} adopted sets of valid inequalities to improve the performance~\cite{schermer2019matheuristic}. Kitjacharoenchai \emph{et al.} developed a solution for the problem to address the limitations of the drone-launch and delivery time~\cite{kitjacharoenchai2019multiple}. Murray \emph{et al.} proposed a solution for an arbitrary number of heterogeneous drones for a truck with specific emphasis on real-world issues~\cite{murray2020multiple}. Compared to existing approaches designed for the limited number of human-driven trucks starting at a fixed location, our work aims to address a new challenge for integrating a large number of available AVs focusing on the optimal selection and assignment of AVs to customer locations to maximize profits for both the logistics company and AV owners.



\section{System Model}
\label{sec:system_model}

There are $N_{v}$ available AVs, denoted by a set $V = \{v_1, v_2, ..., v_{N_v}\}$. The 2D location of an AV $v$ is denoted by $p_v = (x_v, y_v)$. There are three different types of AVs, \emph{i.e.,} SUVs (sport utility vehicles), trucks, and passenger vehicles. The operation cost of an AV differs depending on whether it is in motion, \emph{i.e.,} it is denoted by $c_v^{M} \in \mathbb{R}$ when it is in motion and $c_v^{S} \in \mathbb{R}$ when it is stationary. It is also assumed that the operation cost differs depending on the type of an AV. A different type of an AV has a different load capacity denoted by $q_v \in \mathbb{N}$, which is basically the number of parcels that can be loaded on the vehicle. Each AV is available only for a certain duration of time denoted by $\tau_v^{A}$, specified by the owner of the AV. An AV should finish delivery before $\tau_v^{A}$ is expired. Additionally, the remaining fuel (or electricity for electric vehicles) of an AV $v$ is denoted by $f_v^{A}$, and the fuel consumption rate of an AV $v$ for the mobile and stationary modes are denoted by $f_v^{M}$ and $f_v^{S}$, respectively.

\begin{table}[t]
\center
\label{table:notations}
\caption{The List of notations}
\begin{tabular}{ l|l }
  \hlinewd{1.5pt}
  Symbol &  Description\\
  \hline
  $V = \{v_1, v_2, ..., v_{N_v}\}$ & \pbox{4.5cm}{Available AVs} \\ \hline
  $c_v^{M}$ & \pbox{4.5cm}{Operation cost of $v \in V$ (mobile)} \\ \hline
  $c_v^{S}$ & \pbox{4.5cm}{Operation cost of $v \in V$ (stationary)} \\ \hline
  $q_v$ & \pbox{4.5cm}{Load capacity of $v \in V$} \\ \hline
  $\tau_v^{A}$ & \pbox{4.5cm}{Available time for $v \in V$} \\ \hline
  $f_v^{M}$ & \pbox{4.5cm}{Fuel consumption rate of $v \in V$ (mobile)} \\ \hline
  $f_v^{S}$ & \pbox{4.5cm}{Fuel consumption rate of $v \in V$ (stationary)} \\ \hline
  $L = \{1, 2, ..., l_{N_l}\}$ & \pbox{4.5cm}{Customer locations} \\ \hline
  $G = \{g_1, g_2, ..., g_{N_g}\}$ & \pbox{4.5cm}{Groups (each group consists of customer locations)} \\ \hline
  $p_{f}$ & \pbox{4.5cm}{Location of the distribution facility} \\ \hline
  $p_{g}^{W}$ & \pbox{4.5cm}{Waiting location for group $g \in G$} \\ \hline
  $\matr{M}^{LG}$ & \pbox{4.5cm}{Matrix defining how customer locations are mapped to groups} \\ \hline
  $\tau_g^{D}$ & \pbox{4.5cm}{Delivery completion time for group $g \in G$} \\ \hline
  $\matr{M}^{VG}$ & \pbox{4.5cm}{Matrix defining which AVs cover which groups} \\ \hline
  $b$ & \pbox{4.5cm}{Available budget} \\ \hline
  \hlinewd{1.5pt}
\end{tabular}
\end{table}

There are $N_l$ customer locations denoted by $L = \{1, 2, ..., l_{N_l}\}$. Parcels are delivered to these customer locations from a distribution center located at $p_{f}$. We assume that the delivery area is divided into zones as practiced by many logistics companies~\cite{jung2006integration}. These customer locations are organized into $N_g$ groups denoted by $G = \{g_1, g_2, ..., g_{N_g}\}$ where each group corresponds to a zone. More specifically, the customer locations of a group $g \in G$ are covered by drones departing from a vehicle parked at a specific location corresponding the group, called the waiting location denoted by $p_{g}^{W}$. The mapping between customer locations and groups is defined as a matrix $\matr{M}^{LG}$. For instance, $\matr{M}^{LG}_{i,j} = 1$ if the customer location $j \in L$ belongs to group $i \in G$. In particular, the expected delivery completion time for each group $g \in G$ denoted by $\tau_g^{D}$ can be pre-computed since the number and position of the customer locations in each group are known. Table II summarizes the notations used throughout this paper.

\section{Last-Mile Delivery Using Autonomous Vehicles with Drones}
\label{sec:problem_definition}

A-VRPD is to select a set of participating AVs denoted by $\overline{V}$ from available AVs $V$ ($\overline{V} \subseteq V$), where each participating vehicle $v \in \overline{V}$ is assigned to serve one or more groups in $G$ in a specific order such that the total cost for delivering parcels to all customers is minimized. A-VRPD is solved in two stages: precomputing traveling distances and times taking into account the current traffic conditions (Subsection~\ref{sec:dist_and_time}) and solving an ILP based on the precomputed distances and times (Subsection~\ref{sec:opt_delivery_scheduling}).

\subsection{Precomputation of Distance and Time Segments}
\label{sec:dist_and_time}

The total cost to cover all customer locations depends on the total distance $d_v^{T}$ that each participating AV $v$ travels and the total amount of time $\tau_v^{T}$ it takes for each trip. Thus, the optimal solution for A-VRPD can be obtained faster by facilitating the computation of $d_v^{T}$ and $\tau_v^{T}$. More specifically, we find that $d_v^{T}$ and $\tau_v^{T}$ can be calculated more quickly since the distance and time segments constituting $d_v^{T}$ and $\tau_v^{T}$ can be pre-computed as the locations of participating AVs, the distribution facility, and the waiting location of each group are known, as well as the current traffic conditions.

Let's denote the length of a route between two 2D locations $i$ and $j$ by $d_{i,j}$, which can be obtained using a navigation tool such as Google Map API. Also, denote by $\tau_{i,j}$ the travel time between two 2D locations $i$ and $j$, which can also be estimated using a navigation tool. We compute all distance segments between (1) the distribution facility and a waiting location, denoted by $d_{f,w_1}$, (2) two waiting locations, denoted by $d_{w_i, w_j}$, and (3) the distribution facility and the vehicle's initial location, denoted by $d_{f,v}$ (or $d_{v,f}$). The total distance $d_v^{T}$ for AV $v$ to visit a sequence of waiting locations $\{w_1, ..., w_n\}$ can then be computed using the pre-computed distance segments as follows.

\begin{displaymath}
d_v^{T} = d_{v,f} + d_{f,w_1} + \sum\limits_{k = 1}^{n-1} d_{w_k,w_{k+1}} + d_{w_n,f} + d_{f,v},
\end{displaymath}

\noindent which indicates that an AV $v$ first moves to the distribution facility $f$ to get loaded with parcels and drones; and then, it visits a sequence of waiting locations $\{w_1, ..., w_n\}$ to serve customers; once it completes visiting all waiting locations, it returns to the distribution facility and moves back to its initial location.

Similarly, the total travel time $\tau_v^{T}$ can be calculated based on the precomputed time segments as the following.

\begin{displaymath}
\tau_v^{T} = \tau_{v,f} + \tau_v^{L} + \tau_{f,w_1} + \sum\limits_{k = 1}^{n-1} \tau_{w_k, w_{k+1}} + \tau_{w_n, f} + \tau_{f,v} + \tau_v^{W}.
\end{displaymath}

\noindent Here $\tau_v^{L}$ is the amount of time required to load parcels and drones on AV $v$, and $\tau_v^{W}$ is the total amount of time AV $v$ has to wait for its drones to complete delivery. Consequently, the computation of $d_v^{T}$ and $\tau_v^{T}$ for AV $v$ can be expedited by using precomputed distance/time segments, leading to accelerated calculation of the optimal solution. Another salient aspect of the proposed pre-computation method is that we take into account the current traffic conditions to calculate the total cost, \emph{i.e.,} the distance and time segments are updated accounting for the current traffic conditions and used for obtaining new optimal solutions.


\subsection{ILP Formulation}
\label{sec:opt_delivery_scheduling}

A-VRPD is formulated as an integer linear program (ILP). The objective function is to minimize the total operation cost, which is the sum of the costs for each participating AV. The cost for each participating AV $v \in \overline{V}$ is calculated based on the total distance it has traveled ($d_v^{T}$) and the total amount of time it has waited for its drones to complete delivery ($\tau_v^{W}$), \emph{i.e.,} the cost for each AV $v$ is $c_v^{M} \cdot d_v^{T} + c_v^{S} \cdot \tau_v^{W}$. Thus, the total cost for all participating AVs is $\sum\limits_{v \in \overline{V}} (c_v^{M} \cdot d_v^{T} + c_v^{S} \cdot \tau_v^{W})$. The ILP is to find a matrix $\matr{M}^{VG}$, which defines the mapping between AVs and customer groups, and a sequence of groups to be covered by each AV, to minimize the total cost. As such, the ILP is defined as follows.

\begin{displaymath}
\begin{aligned}
\argminA_{\matr{M}^{VG}} \quad & \sum\limits_{v \in \overline{V}} (c_v^{M} \cdot d_v^{T} + c_v^{S} \cdot \tau_v^{W})\\
\textrm{s.t.} \quad & \sum\limits_{i \in V}\matr{M}^{VG}_{i,j} = 1, \forall j \in G& (1)\\
& \sum\limits_{i \in \overline{V}} (c_i^{M} \cdot d_i^{T} + c_i^{S} \cdot \tau_i^{W}) \le b & (2)\\
& \sum\limits_{j \in G} (\matr{M}^{VG}_{i,j} \cdot \sum\limits_{k \in L}\matr{M}^{LG}_{j,k}) \le q_i, \forall i \in V & (3)\\
& (f_v^{M} \cdot d_i^{T} + f_v^{S} \cdot \tau_i^{W}) \le f_v^{A}, \forall i \in V & (4)\\
& \tau_i^{T} \le \tau_i^{A}, \forall i \in V & (5)\\
& \tau_i^{W} = \sum\limits_{j \in G} \matr{M}_{ij}^{VG} \cdot  \tau_j^{D}, \forall i \in V & (6)\\
\end{aligned}
\end{displaymath}

Constraint (1) ensures that each group is covered by only one AV, and all groups are covered. Constraint (2) indicates that the total cost should be smaller than the available budget $b$. Constraint (3) dictates that the number of customer locations covered by an AV should be smaller than the AV's capacity. Recall that $\matr{M}^{VG}_{i,j} = 1$, if group $j \in G$ is covered by vehicle $i \in V$. Thus, if we multiply $\matr{M}^{VG}_{i,j}$ by $\sum\limits_{k \in L}\matr{M}^{LG}_{j,k}$ (\emph{i.e.,} the total number of customer locations of group $j$), we obtain the total number of customer locations of the group covered by an AV $i \in V$. If we repeat this computation for all groups, we get the total number of customer locations covered by an AV $v$, which should be smaller than its capacity. Constraint (4) is used to ensure that each AV has enough gas to serve all customer locations assigned to it. Constraint (5) enforces that the total operation time of an AV does not exceed the available time of the AV. This means that an AV should return to its original location before its available time is expired. Finally, Constraint (6) is an equality constraint that defines the total amount of time that an AV has waited for its drones to complete delivery.

To evaluate a solution for the ILP, which is given in the form of a matrix $\matr{M}^{VG}$, the solver requires one additional information: the sequence of groups visited by each participating AV. More specifically, this sequence of groups is used by the solver to compute the total distance traveled $d_v^{T}$ and the total travel time $\tau_v^{T}$. Since $d_v^{T}$ has the greatest impact on the total cost, the sequence of groups can be determined such that $d_v^{T}$ is minimized. The problem of determining the sequence of groups such that $d_v^{T}$ is minimized is exactly the traveling salesman problem (TSP), which is in NP-Hard (as such, A-VRPD is in NP-Hard). Fortunately, considering the relatively small capacity of AVs as well as the fact that each group consists of multiple customer locations, AVs are supposed to cover only a small number of groups. Consequently, we find that calculating the optimal sequence of groups using a standard solver for TSP does not significantly degrade the computational delay for solving A-VRPD (Section~\ref{subsec:save_money}).

\section{Greedy Approach}
\label{sec:greedy_approach}

We present a greedy heuristic algorithm to solve A-VRPD (which is in NP-Hard) much more quickly. The basic idea is to keep greedily selecting a participating AV that serves the greatest number of customer locations with a minimum cost until all customer locations are covered, or until the available budget $b$ is exhausted. The total operation cost $c^{f}$ is defined as the sum of all `per-vehicle' costs for selected AVs.

\begin{wrapfigure}{r}{0.65\columnwidth}
\vspace{-10pt}
  \begin{center}
    \includegraphics[width=\linewidth]{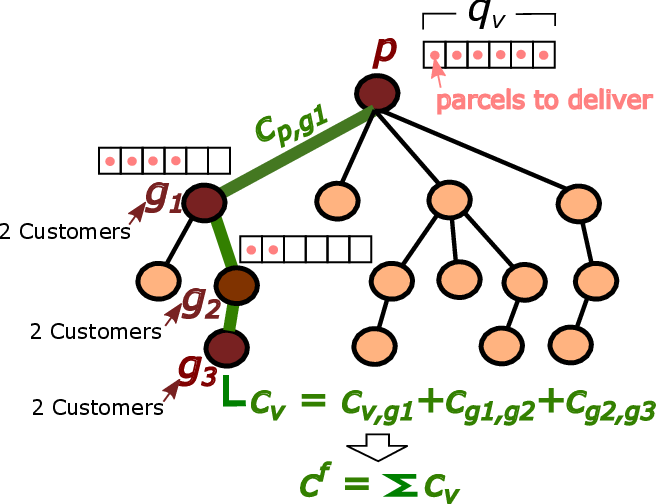}
    \caption{An illustration of a tree structure constructed by the greedy algorithm. \label{fig:tree_based_approach}}
  \end{center}
  \vspace{-10pt}
\end{wrapfigure}


The greedy algorithm constructs a tree structure to calculate the per-vehicle cost $c_v$ for AV $v$. We first explain how the tree is built and describe how the per-vehicle cost is calculated based on the tree. Consider Fig.~\ref{fig:tree_based_approach} for an example. The root of the tree represents the distribution center. Each node in the first level of the tree represents a group that can be covered by an AV at the distribution center, \emph{i.e.,} the number of customer locations of the group is smaller than the number of parcels loaded on the AV. As shown in Fig.~\ref{fig:tree_based_approach}, the capacity $q_v$ of AV $v$ is six which means that at most six parcels can be loaded. With this capacity, the four nodes in the first level of the tree mean the groups that can be served by AV $v$.

Now, assume that one of the nodes in the first level is covered by AV $v$. For example, in Fig.~\ref{fig:tree_based_approach}, group $g_1$ having 2 customers is highlighted to indicate that $g_1$ is covered by the vehicle. Since the two customers of $g_1$ have been served, there are four parcels left on the vehicle. The algorithm then repeats the above-mentioned process to find the nodes in the second level, and so on, until all parcels on AV $v$ are delivered to customers, thereby completing construction of the tree. The algorithm also ensures that an AV that does not have enough fuel is not selected.


\begin{algorithm}
  \label{algorithm1}
  \caption{The Greedy Algorithm.}
  \textbf{Input:}  $V$, $G$, $c_v^{M}$, $c_v^{S}$, $Q_v$, $\tau_{g}^{D}$, $b$, $f_v^{M}, f_v^{S}, f_v^{A}$, \;
  $d_{i,j}$ (pre-computed distance segments)\;
  \textbf{Output:} $c^{f}$ (final cost)\;
  $c^{f} \leftarrow 0$\;
  \While {$|G| > 0$ and $c^{t} < b$} {
    \For {each vehicle $v \in V$} {
      $t \leftarrow$ build\_tree(v)\;
      $c_v \leftarrow \mbox{compute\_cost(t, v)}$\;
      $C_V \leftarrow C_V \cup c_v$\;
    }
    // Find $v' \in V$ with the minimum cost using $C_V$\;
    $v' \leftarrow$ find\_min\_cost\_vehicle($C_V$)\;
    $G \leftarrow G \setminus G_{v'}$ //$G_{v'}$ is the groups covered by $v'$\;
    $c^{f} \leftarrow c^{f} + c_{v'}$\;
    $C_V \leftarrow \emptyset$\;
  }
\end{algorithm}

Once the tree is constructed for AV $v$, using the tree, the greedy algorithm calculates the per-vehicle cost $c_v$. The per-vehicle cost is calculated as the sum of the group costs for each group that is covered by the AV. More specifically, the group cost $c_{p,g}$ to serve a group $g$ from the current location of the AV denoted by $p$ is defined as $(d_{p,g} c_v^{M} + \tau_{g}^{D} c_v^{S}) / |g|$ where $|g|$ is the number of customer locations of group $g$. The reason why we divide by $|g|$ is to ensure that serving a group having more customers is more cost-effective assuming the same traveling distance to the group. The group cost for each group is represented as the weight of an edge of the tree as shown in Fig.~\ref{fig:tree_based_approach}; thus, if we consider the sum of the weights of edges on a simple path from the root to a terminal node as the cost for the simple path, the per-vehicle cost $c_v$ is calculated as the smallest cost of all such simple paths. Incidentally, the simple path with the smallest per-vehicle cost represents the sequence of the groups visited by the AV. The greedy algorithm repeats this process of computing per-vehicle cost $c_v$ for all AVs and selects an AV $v'$ with the smallest per-vehicle cost $c_v'$ as a participating AV. Once such a participating AV is found, the groups covered by the selected AV are excluded as they have been already covered; and then, the per-vehicle cost $c_v'$ is added to the final cost $c^{f}$. The greedy algorithm keeps finding a participating AV in this greedy manner until all customer locations are covered or the available budget is exhausted, obtaining the final cost $c^{f}$. Algorithm 1 summarizes the operation of the greedy algorithm.

It is worth mentioning that the same AV can be selected by the algorithm, \emph{e.g.,} in case there are not enough AVs. However, when the same AV is selected, the cost for returning to the distribution center is added because the AV has to return to the distribution center to reload parcels. We also note that, in computing the group cost $c_{p,g}$, if $g$ is the first group to serve, the cost for moving an AV from the AV's original position to the distribution center is added to the partial cost; similarly, if $g$ is the last group that AV $v$ serves, the cost for moving the AV from $g$ to the original position of the AV is added unless the vehicle is sent to the distribution center for getting reloaded.


\section{Simulation Results}
\label{sec:evaluation}

The MATLAB optimization toolbox~\cite{matlab} is used to solve A-VRPD and implement the greedy algorithm. A PC equipped with a Intel Core i7-9750H processor with 16GB RAM running on Windows 10 is used for this simulation study. Ten random delivery scenarios are created where 500 customer locations are randomly selected from the city of Memphis in the US. These 500 customer locations are distributed into 80 groups. Different types of AVs are used including compact sedans (type 1), SUVs (type 2), and pickup trucks (type 3). The default number of AVs available for delivery is set to 50. All AVs are uniformly distributed in the region.


\begin{table}[h]
\center
\caption {The default input parameters used for simulation}
  \begin{tabular}{ l | l }
  \hlinewd{1.5pt}
   Parameters & Values (type 1, type 2, type 3)\\ \hline
  $b$ & \$1,200  \\ \hline
    $c_v^{M}$ & \$0.1, \$0.12, \$0.15  \\ \hline
    $c_v^{S}$ & \$0.00013, \$0.00033, \$0.00071 \\ \hline
    $f_v^{A}$ & 13 gal, 15 gal, 23 gal  \\ \hline
    $q_v$ & 7, 9, 14  \\
  \hlinewd{1.5pt}
\end{tabular}
\end{table}

A main metric used in this simulation study is the amount of profits both for the logistics company and AV owners. To calculate profits for participating AV owners, the 50:50 profit model is used that allows the delivery company and participating AVs to equally share the profits. Another important metric is the ``delivery finish time'' that measures how long it takes to complete to serve all customers. We measure these two metrics by varying delivery scenarios and the number of available AVs. Table III summarizes the default parameters. The fuel costs for different types of vehicles are determined based on real-world data~\cite{fuelcost}. In particular, the budget $b$ is determined according to~\cite{amazon2}; a truck driver makes approximately \$24 per hour and covers about 100 customers a day~\cite{allen2018understanding, amazon1}. Therefore, the per-customer cost is roughly \$2.4. Since there are 500 customers to serve, the budget $b$ is set to \$1,200.

\subsection{Profits Analysis}
\label{subsec:save_money}

We analyze the performance of the greedy and optimal algorithms by measuring the profits both for the delivery company and AV owners. Here, the profits are defined as the budget $b$ minus the total cost to serve all customers. The greedy and optimal algorithms are compared with two different versions of base algorithms. The first one, denoted by Base-1, is implemented based on a TSP-D algorithm~\cite{bouman2018dynamic}; in this algorithm, an AV is randomly selected from a pool of available AVs; the selected AV serves customers based on a path generated by the underlying TSP-D algorithm; the Base-1 algorithm keeps selecting AVs randomly and allows them to cover customers based on the TSP-D algorithm until all customers are served. The second one, denoted by Base-n, is similar to Base-1, but the difference is that $n$ vehicles are randomly selected from the pool of available AVs; as such Base-n is implemented using a more recent VRP-D algorithm~\cite{tamke2021branch}.

\begin{figure}[!htbp]
\begin{minipage}[b]{0.48\columnwidth}
\centering
  \includegraphics[width=\linewidth]{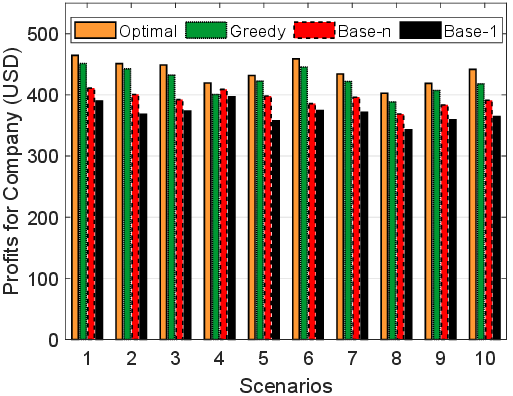}
\caption {Profits for the delivery company in different scenarios.}
\label{fig:money_saved}
\end{minipage}
\hspace{1mm}
\begin{minipage}[b]{0.47\columnwidth}
\centering
\includegraphics[width=\linewidth]{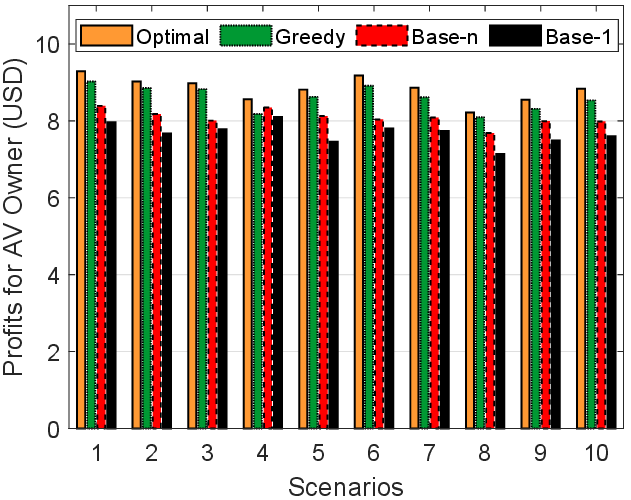}
\caption {Profits for an AV owner in different scenarios.}
\label{fig:money_saved_per_vehicle}
\end{minipage}
\end{figure}



The profits for the delivery company are depicted in Fig.~\ref{fig:money_saved}. The average profits for the greedy and optimal algorithms are \$423.1 and \$437.2, respectively, demonstrating that the optimal algorithm achieves 3.3\% higher profits compared with the greedy algorithm. Such a small gap is noteworthy considering the significant difference in the computation times of the two algorithms; the average computation time for the greedy and optimal algorithms under the same simulation settings are 3.8 sec and 176.2 sec, respectively. We also compare the profits of the two algorithms with that of the base algorithms. The greedy algorithm achieves higher profits by 14.4\% and 7.5\% compared with Base-1 and Base-n algorithms, respectively. The reason for the higher profits is attributed to the fact that the proposed algorithms are designed to effectively select AVs to optimize profits and take into account the current traffic conditions to create paths for AVs.

The profits for each AV owner are also measured and displayed in Fig.~\ref{fig:money_saved_per_vehicle}. The results demonstrate that average profits for each AV owner are \$8.6 and \$8.8 when the greedy and optimal algorithms are used, respectively. We observe that the greedy algorithm, in particular, achieves 12.2\% and 6.5\% higher profits, respectively, compared with that of Base-1 and Base-n algorithms, respectively. The results are interesting since they represent only daily profits (about \$250 per month) after the fuel cost, involving no human labor at all. Furthermore, if a different profit model is used other than the 50:50 model, \emph{e.g.,} to attract the participation of more AV owners, the profits for AV owners can be increased.

\subsection{Effect of Number of Available AVs}
\label{subsec:number_of_AVs}

\begin{figure}[!htbp]
\begin{minipage}[b]{0.482\columnwidth}
\centering
\includegraphics[width=\linewidth]{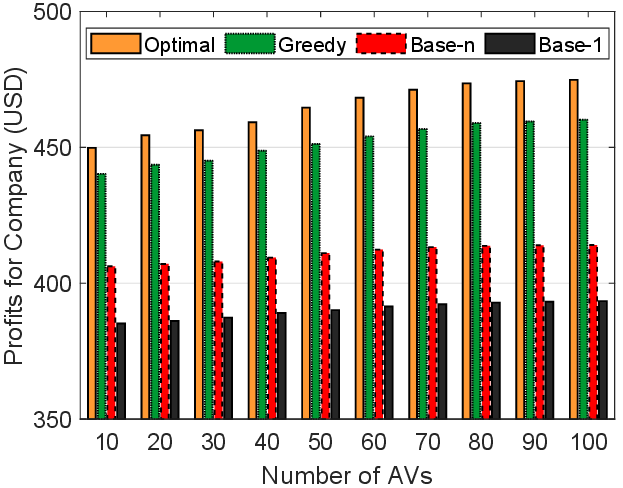}
\caption {Profits for the delivery company with varying numbers of AVs.}
\label{fig:money_saved_per_num_av}
\end{minipage}
\hspace{1mm}
\begin{minipage}[b]{0.468\columnwidth}
\centering
\includegraphics[width=\linewidth]{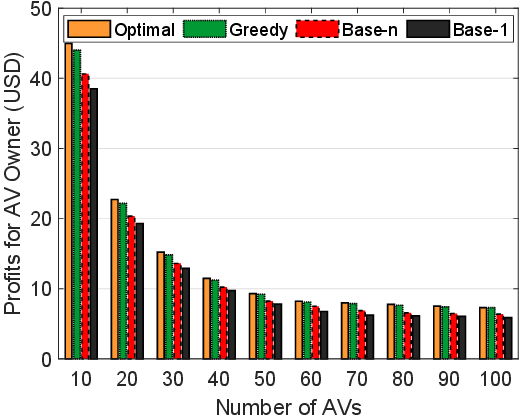}
\caption {Profits for participating AV owners with varying numbers of AVs..}
\label{fig:money_earned_per_num_av}
\end{minipage}
\end{figure}

In this section, we analyze the effect of the number of available AVs on the profits for the delivery company and AV owners. More specifically, we run the greedy, optimal, Base-1, and Base-n algorithms and measure the profits by varying the number of available AVs from 10 to 100 with an interval of 10 (default = 50 AVs). Fig~\ref{fig:money_saved_per_num_av} depicts the profits for the delivery company. It is observed that the profits generally increase as the number of available AVs increases regardless of the algorithms. The reason for lower profits when there are not enough AVs is due to the additional costs for AVs to return to the distribution center to reload parcels. In contrast, when there are sufficient AVs, all customers are served without having to make some AVs to return to the distribution center to get reloaded.

We also measure the profits for AV owners by varying the number of available AVs. Fig.~\ref{fig:money_earned_per_num_av} depicts the results. As expected, we observe that AV owners make higher profits when there is not enough AVs despite the higher fuel cost used to return to the distribution center to reload parcels. On the other hand, it is observed that the profits for them decrease significantly as the number of AVs increases. An interesting observation is that the decreasing trend flattens out as the number of available AVs increases indicating that the algorithms do not recruit AVs more than they need to maximize the profits.

\subsection{Delivery Finish Time}
\label{subsec:delivery_completion_time}

\begin{wrapfigure}{r}{0.5\columnwidth}
\vspace{-45pt}
  \begin{center}
    \includegraphics[width=\linewidth]{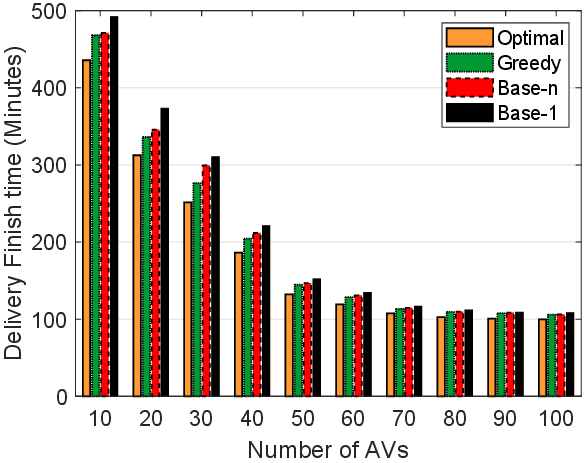}
    \caption{The delivery finish time for different numbers of AVs. \label{fig:delivery_time}}
  \end{center}
  \vspace{-15pt}
\end{wrapfigure}


We have demonstrated that AV owners potentially make higher profits when there are relatively small number of available AVs. Conversely, when there is not enough AVs, the delivery finish time is expected to increase since AVs will have to return to the distribution center to reload parcels. In this section, we analyze the delivery finish time by varying the number of available AVs. Fig.~\ref{fig:delivery_time} depicts the results demonstrating that the delivery finish time is significantly higher when there are only 10 AVs in comparison with the scenario with 100 available AVs; more precisely, the delivery finish time for 10 available AVs is 353\% and 345\% higher when the optimal and greedy algorithms are used, compared with those for 100 AVs, respectively. It is also observed that as more AVs become available, the delivery finish time significantly decreases. The results indicate that having a large number of available AVs is more profitable for the delivery company not only because of the smaller delivery finish time but also due to the higher expected profits. Another interesting observation is that the greedy algorithm has 6.3\% and 2.5\% smaller delivery finish time compared with the Base-1 and Base-n algorithms, respectively. The reason is that the greedy algorithm strategically selects AVs to minimize the cost.

\section{Conclusion}
\label{sec:conclusion}

We have presented an ILP formulation of A-VRPD and developed a greedy heuristic solution to solve the problem. The proposed algorithm efficiently selects AVs from a pool of available AVs and optimally assigns them to customer locations to maximize profits not only for the logistics company but also for AV owners. Through extensive simulations performed with realistic delivery scenarios, fuel costs, and salaries for truck drivers, we demonstrated that the proposed greedy algorithm achieved 14.4\% (12.2\%) and 7.5\% (6.5\%) higher profits for the delivery company and AV owners, respectively, compared with traditional TSP-D algorithm-based (VRP-D algorithm-based) approaches. We claim that this work is expected to open the door for new research on methodologies for fully automating drone-based last-mile delivery by replacing human-driven trucks with autonomous vehicles.

\bibliographystyle{IEEEtran}
\bibliography{mybibfile}

\end{document}